\documentclass[twoside]{ilcws07}
\usepackage[latin1]{inputenc}
\usepackage[dvips]{graphicx,epsfig,color}
\usepackage{wrapfig,rotating}
\usepackage{amssymb,amsmath,array}

\begin{document}
\title{
%%%%   Paper title goes here  %%%%%%%%%%%%%%
MarlinTPC: A Marlin based common TPC software framework for the LC-TPC collaboration} %% 
%***********************************************************************
% AUTHORS INFORMATION AREA
%***********************************************************************
\author{Jason Abernathy$^1$, Klaus Dehmelt$^2$, Ralf Diener$^2$, Jim
  Hunt$^3$,\\ Matthias Enno Janssen$^2$, Martin Killenberg$^4$,
  Thorsten Krautscheid$^4$,\\ Astrid M\"unnich$^5$, Martin
  Ummenhofer$^4$, Adrian Vogel$^2$ and Peter Wienemann$^4$
% Optional short acknowledgment: remove next line if non-needed
%\thanks{This is an optional funding source acknowledgment.}
% DO NOT MODIFY THE FOLLOWING '\vspace' ARGUMENT
\vspace{.3cm}\\
% Addresses and institutions (remove "1- " in case of a single institution)
1- University of Victoria - Department of Physics and Astronomy \\
PO Box 3055, V8W 3P6 Victoria BC - Canada
%% Remove the next three lines in case of a single institution
\vspace{.1cm}\\
2- Deutsches Elektronen-Synchrotron\\
Notkestr. 85, 22607 Hamburg - Germany
%% Remove the next three lines in case of a single institution
\vspace{.1cm}\\
3- Cornell University, Wilson Laboratory\\
Ithaca, NY 14853 - USA
%% Remove the next three lines in case of a single institution
\vspace{.1cm}\\
4- University of Bonn, Department of Physics\\
Nussallee 12, 53115 Bonn - Germany
%% Remove the next three lines in case of a single institution
\vspace{.1cm}\\
5- RWTH Aachen, III.~Physikalisches Institut \\
Physikzentrum, 52056 Aachen - Germany\\
}
%%***********************************************************************
% END OF AUTHORS INFORMATION AREA
%***********************************************************************

\maketitle

\begin{abstract}
  We describe the goals and present functionality of MarlinTPC, a
  common software framework for the LC-TPC collaboration based on
  LCIO, Marlin and other ilcsoft tools.
\end{abstract}

\section{Introduction}

Three of the four available designs for detectors at the International
Linear Collider (ILC) envisage a large time projection chamber (TPC)
as main tracking device. It is the task of the LC-TPC collaboration to
perform the necessary R\&D to be able to fulfil the TPC performance
requirements derived from the ILC physics goals \cite{ref:RDR}.

In the course of LC-TPC activities a rather large variety of
simulation, analysis and reconstruction software packages has been
developed.  The usage of these packages ranges from studying TPC
performence as part of an overall detector for different detector
layouts and background conditions, optimising prototype designs,
reconstructing and analysing cosmics and testbeam data from various
small prototypes using different readout technologies to full detector
physics analyses.  Most of these individual software packages have
become rather sophisticated in their particular field of application.
Valuable experience was collected during usage and development of the
software partly leading to novel techniques to cope with new
challenges encountered in TPCs with new amplification or readout
systems.  The drawback of this specialisation is that the software
often works smoothly only for particular applications or TPC setups.
Exchanging code between different packages or analysing data from
different sources with the same program can be very time consuming and
errorprone since the different programmes do not use commonly accepted
interfaces and conventions. The goal of MarlinTPC is to overcome these
drawbacks.

\section{Design considerations}

In view of the converging hardware efforts like the planned common
large prototype, to improve the mutual understanding of results and to
avoid further double work, it seems natural to converge on the
software tools. In June 2006 an initiative was started to take the
first step towards this direction. Representatives of six LC-TPC
member institutes met at DESY to find an agreement on common software
standards. It was decided to transfer the existing algorithms to a
new, commonly used framework, called MarlinTPC \cite{ref:MarlinTPC},
building on top of LCIO \cite{ref:LCIO}, the {\it de facto} standard
data format for ILC related work, and the accompanying ilcsoft tools
\cite{ref:ILCSoft}.  This choice is motivated by the possibility to
profit from general ILC software developments and by the fact that
ilcsoft tools are already used by many other subdetector, simulation
and physics analysis software projects.  In particular Marlin
\cite{ref:Marlin} was chosen as analysis and reconstruction framework.
Its modularity and well defined interfaces between its modules, called
processors, ensure that different developers can work on different
processors in parallel without interference and to plug'n'play with
processors to easily try out e.~g.~different algorithms.

Further important pillars of MarlinTPC are the Linear Collider
Conditions Data (LCCD) toolkit \cite{ref:LCCD} and GEAR (Geometry API
for Reconstruction) \cite{ref:GEAR}. LCCD allows writing and reading
of conditions data describing the detector status as function of time.
It allows to tag data sets for later easy reference and to request
data valid at a particular point in time. GEAR allows to access
geometry information needed for the simulation, digitisation and
reconstruction of events. Thereby it is made sure that consistent
geometry information is used throughout MarlinTPC.

\section{Reconstruction}

In the beginning the focus was laid on the development of
reconstruction processors. Therefore the reconstruction code is
currently the most evolved part of MarlinTPC. It will be briefly
described in this section.

\subsection{Processor chain}

The event data model (EDM) for the reconstruction is provided by the
LCIO classes {\tt TrackerRawData}, {\tt TrackerData}, {\tt
TrackerPulse}, {\tt TrackerHit} and {\tt Track}. These data structures
represent well-defined interfaces between the different reconstruction
steps. Every processor retrieves one or several input collections
which contain the input data (e.~g.~{\tt TrackerData} objects),
processes them (e.~g.~applies a pulse finding algorithm) and finally
provides output collections containing the results of the applied
algorithm (e.~g.~{\tt TrackerPulse} objects).  The output collections
in turn can be read in by subsequent processors to further process the
event data.

The present processor chain for the MarlinTPC reconstruction is shown
in Table \ref{tab:RecoProcessors}. Additional correction processors
e.~g.~to correct for electric or magnetic field inhomogeneities or
gain fluctations can be easily added in the appropiate places if
needed.

\begin{table}
{\footnotesize
\begin{tabular}{l c r}
Data structure & Processor name & input/output collection name \\ \hline
{\tt TrackerRawData} & & TPCRawData\\
                     & TrackerRawDataToDataConverterProcessor & \\
{\tt TrackerData}    & & TPCConvertedRawData \\
                     & PedestalSubtractorProcessor & \\
%                     & ChannelByChannelCorrectorProcessor & \\
%                     & LinearityCorrectorProcessor & \\
                     & TimeShiftCorrectorProcessor & \\
{\tt TrackerData}    & & TPCData\\
                     & PulseFinderProcessor & \\
                     & ChannelMapperProcessor & \\
%                     & GainCorrectorProcessor & \\
                     & CountsToPrimaryElectronsConverterProcessor & \\
{\tt TrackerPulse}   & & TPCPulses \\
                     & HitTrackFinderTopoProcessor & \\
{\tt TrackerHit}, {\tt Track}    & & TPCHits, TPCTrackCandidates \\
                     & TrackSeederProcessor & \\
{\tt Track}          & & TPCSeedTracks \\
                     & TrackFitterLikelihoodProcessor & \\
{\tt Track}          & & TPCTracks
\end{tabular}
}
\caption{Present MarlinTPC reconstruction processors}
\label{tab:RecoProcessors}
\end{table}

\subsection{Present functionality}

This section briefly describes the functionality of the MarlinTPC
reconstruction as it is available in September 2007.

{\tt TrackerRawDataToDataConverterProcessor} converts {\tt
TrackerRawData} objects with integer FADC counts into {\tt
TrackerData} objects with floating point numbers for the channel time
spectrum. This allows for pedestal subtraction by the {\tt
PedestalSubtractorProcessor} since the pedestals are in general
non-integer values. The pedestals are provided by LCCD and are read in using
the Marlin {\tt ConditionsProcessor}.
 Optionally further correction processors can by
applied after the pedestal subtraction.

 The next step is the search
for pulses in the channel time spectra performed by the {\tt
PulseFinderProcessor}. It uses a threshold based method which has an
individual threshold per channel depending on the noise width calculated from
the pedestal. The algorithm is capable to handle signals with positive and
negative polarity as well as zero suppressed data to support a large variety
of possible readout electronics.

The {\tt ChannelMapperProcessor} translates the hardware channel numbers of the
electronics into GEAR pad indices. Afterwards the number of ADC counts per
pulse is converted into primary electrons by the
{\tt CountsToPrimaryElectronsConverterProcessor}, using electronics and gas gain
calibration factors.
The channel map as well as the calibration constants are stored using LCCD.

The {\tt HitTrackFinderTopoProcessor} performs a topological search for
pulses on neighbouring pads. 
Pulses on contiguous pads in one pad row are grouped as hit candidates.
Afterwards 
the hit coordinates are calculated. Contiguous hits in different pad rows
within an adjustable time window are
assembled to make up a track candidate. The algorithm does not make any
assumption on the trajectory and thus works for straight tracks like
cosmic muons in a prototype as well as for a low energetic particle curling up
in the TPC, 
producing a helix with a changing curvature due to energy loss. It is
implemented for the two available GEAR pad geometries (rectangular and
circular pad row layout).
An estimate for the track parameters is calculated analytically by the {\tt
  TrackSeederProcessor} to have a good starting point for the succeeding fit.

The {\tt TrackFitterLikelihoodProcessor} determines the track
parameters by maximising the global likelihood for observing the
measured charge distribution on all pads associated with the track. It
has been shown to give better results than just minimising the mean
squared distance of the reconstructed hits to the track using a
$\chi^2$ fit \cite{ref:LikehoodChi2}.

For bookkeeping every processor writes all its processor parameters like cuts
etc.\ to the LCIO run header, as well as the subversion software revision of
the code which was used to process the data.

So far the code has just been validated with toy data. Thus the
performance of the available processors still needs to be checked with
real data or more realistic Monte Carlo samples.

\section{Simulation, digitisation and analysis}

Very recently sizable development work on simulation, digitisation and
analysis code has started.

The simulation part of MarlinTPC started off with the package
TPCGEMSimulation from \cite{ref:TPCGEMSimulation}. It simulates
primary ionisation using a parametrisation of HEED \cite{ref:HEED}
output, drifting of the primary electrons to the endplates,
amplification in a triple-GEM gas amplification system and electronics
shaping. Optionally an ion backdrift processor can be added. The
supported gases are Ar-CH$_4$ (95-5), Ar-CH$_4$ (90-10) and
Ar-CH$_4$-CO$_2$ (93-5-2). The original package (which received
geometry information as processor parameters) has been modified such
that all geometry information is retrieved from GEAR now. At present
it is not possible to simulate other gases or amplification systems.
It is planned to separate out the digitisation components of the
package and, in the long run, to make the simulation more generally
usable to be able to simulate other gases or amplification systems.

Work has started on a set of digitisation processors which can be used
to digitise both the output of detailed gas detector simulations
generating individual ionisation clusters and the output of Geant4
\cite{ref:Geant4} simulations providing rather large energy deposits
instead of individual electron-ion pairs. This also includes the
development of processors producing pile-up due to the large drift
time of electrons (compared to the bunch crossing rate) and background
from $\text{e}^+\text{e}^-$ pairs from the fusion of beamstrahlung
photons.

For analysis purposes several processors are under development which
provide information according to the recommendations of the first ILC
TPC Analysis Jamboree \cite{ref:Jamboree}. Examples are the residual
distributions, the fraction of 1-pad, 2-pad, 3-pad hits, residuals as
function of the position on the pad, etc.

\section{Conclusions and outlook}

Within its first year the MarlinTPC project has made significant
progress. The most important reconstruction processors are available,
partly using newly developed, more powerful algorithms than what was
available in the old software packages. It is planned to extend the
reconstruction to include algorithms needed to handle data from TPCs
with pixel readout such as Medipix2 or Timepix readout chips.  Efforts
are under way to also get similarly powerful simulation, digitisation
and analysis processors.

MarlinTPC will be finally put to the test
with the completion of the common large TPC prototype. With the advent
of data from this prototype it has to prove its capabilities.

\section{Acknowledgments}

This work is partly funded by the Commission of the European
Communities under the 6$^{\text{th}}$ Framework Programme ``Structuring the
European Research Area'', contract number RII3-026126.

% ****************************************************************************
% BIBLIOGRAPHY AREA
% ****************************************************************************

\begin{footnotesize}
% IF YOU DO NOT USE BIBTEX, USE THE FOLLOWING SAMPLE SCHEME FOR THE REFERENCES
% ----------------------------------------------------------------------------

% ----------------------------------------------------------------------------

% IF YOU USE BIBTEX,
% - DELETE THE TEXT BETWEEN THE TWO ABOVE DASHED LINES
% - UNCOMMENT THE NEXT TWO LINES AND REPLACE 'Name_Of_Your_BibFile'

%\bibliographystyle{unsrt}
%\bibliography{Name_Of_Your_BibFile}
% example of Name_Of_Your_BibFile.bib
% @Article{Turcato:2006ch,
%      author    = "Turcato, M.",
%  collaboration = "ZEUS and H1",
%      title     = "Lepton flavour violation and charmonium physics at HERA",
%      journal   = "Nucl. Phys. Proc. Suppl.",
%      volume    = "162",
%      year      = "2006", 
%      pages     = "283-287",
%      SLACcitation  = "%%CITATION = NUPHZ,162,283;%%"
% }
% 
% @Unpublished{Gogitidze:2007du,
%      author    = "Gogitidze, N.",
%  collaboration = "H1", 
%      title     = "Prompt photons and particle momentum distributions at
%                   HERA", 
%      year      = "2007",
%      note    = "hep-ex/0701033",
%      SLACcitation  = "%%CITATION = HEP-EX 0701033;%%"
% }

\end{footnotesize}

% ****************************************************************************
% END OF BIBLIOGRAPHY AREA
% ****************************************************************************

\end{document}